\documentstyle[]{l-aa}
\input psfig.tex
\begin{document}
\topmargin 25mm 
\thesaurus{11(02.12.2; 12.04.01; 11.09.4; 09.19.1; 13.19.1; 13.19.3) }

\title{Perspectives for detecting cold H$_2$ in outer galactic disks}

\author{Fran\c coise Combes \inst{1} 
\and Daniel Pfenniger \inst{2}}

\institute{ DEMIRM, Observatoire de Paris,
 61 Av. de l'Observatoire, 75\ts014 Paris, France
\and  Observatoire de Gen\`eve, CH-1290 Sauverny, Switzerland}

\offprints{F.~Combes \hfill\break(e-mail: bottaro@obspm.fr)}
\date{Received ???; accepted ???}
\maketitle
\markboth{F. Combes \& D. Pfenniger: 
Detecting cold H$_2$ in outer galactic disks}{}

\begin{abstract}
We review here the main direct or indirect ways to detect the possible
presence of large amounts of cold molecular hydrogen in the outer
parts of disk galaxies, an hypothesis that we have recently developed.
Direct ways range from H$_2$ absorption in the UV domain to detection
of the radio hyperfine structure: the ortho-H$_2$ molecule has an
hyperfine, or ultrafine, structure in its fundamental state, due to
the coupling between the rotation-induced magnetic moment, and the
nuclear spin.  This gives rise to 2 magnetic dipole transitions, at the
wavelengths of 0.5 and $5.5\,\rm km$.  Indirect ways are essentially
the detection of the HD and LiH transitions, and in some environments
like clusters of galaxies, more heavy trace molecules such as CO.  We
discuss from this point of view the recent discovery by COBE/FIRAS of
a very cold Galactic dust component ($4-7\,\rm K$) which could
correspond to a dominating gas mass component of the ISM, 
if interpreted as standard dust emission.
Some of the proposed means could be applied to the well-known
molecular clouds, to bring some new light to the problem of the
H$_2$/CO conversion ratio.

\end{abstract}

\section{Introduction}
Molecular hydrogen has been recognized for a long time to be
potentially the most important (baryonic) element in the Universe
(Zwicky 1959), or at least in the Galaxy (Gould, Gold \& Salpeter
1963) (see also the review by Field et al.~1966 and references
therein).  Since the late sixties the availability of millimetric
radio-telescopes allowed to find substantial amounts of molecular
hydrogen in locations where it can be traced by the CO molecule, the
molecular clouds.  Roughly, the indirect detection of molecular
hydrogen by the CO doubled the known gas content of spirals.

However, in metal poor regions where CO can hardly trace H$_2$, the
question is still open.
Most of the mass in the outer, metal poor parts of a galaxy could be
very cold H$_2$, and remains invisible even today.  The present trend
to increase the conversion factor $X$ between the CO intensity and the
H$_2$ mass at decreasing metallicity goes in this direction
(e.g.~Sakamoto 1996).

The difficulty to detect the molecule H$_2$ comes from its symmetry,
cancelling any electric dipole moment in the ground state.  This
prevents the detection of any radio line in emission, the only ones to
be excited from the ground state in the cold interstellar medium
($T\approx10\,\rm K$).  The first H$_2$ lines to be detected in
emission occur only by quadrupole radiation: they are the pure
rotation ones, the first being the $J=2-0$ at $28\,\mu$ wavelength,
already at $512\,\rm K$ above the fundamental state, the $J=3-1$ at
$17\,\mu$ and $J=4-2$ at $12\,\mu$ (Beck et al.~1979, Parmar et
al.~1994), and $J=2-0$), and $2.12\,\mu$ ($v=1-0$, $J=3-1$),
corresponding to an excitation energy of $\approx7000\,\rm K$. These
lines trace hot H$_2$ gas, for example excited in supersonic shocks,
which corresponds to an insignificant fraction of the H$_2$ mass
(e.g.~Shull \& Beckwith 1982).  Many recent results obtained with the
ISO satellite, with the SWS instrument sensitive
between 2 and 45$\mu$, have revealed the detection of more than a
dozen of these ro-vibrational H$_2$ lines (Timmermann et
al.~1996). These lines are emitted by very dense and warm gas ($T>
700\,\rm K$) in shocks, or even hot gas ($T\sim 11'000\,\rm K$), that
is likely excited non-collisionally through UV pumping (Wright et
al.~1996).  In the violent starburst Arp$\,220$, the H$_2$ emitting
warm gas could represent up to 10\% of the ISM mass (Sturm et
al.~1996).

The only direct information on colder H$_2$ gas has been gained
through far UV absorption in front of stars in the solar neighbourhood
by the Copernicus satellite (e.g.~Spitzer \& Jenkins
1975). Unfortunately, only very diffuse clouds could be traced, since
high H$_2$ column densities are associated with
large extinctions. The density (lower than $1\,\rm cm^{-3}$), the
molecular fraction (0.25) and temperature ($\approx 80\,\rm K$) are
typical of diffuse intercloud material. The dense and much colder
molecular cloud bulk component remains to be explored.

In a recent model, we propose that the dark matter detected around
spiral galaxies through flat rotation curves, is composed for a
substantial part of cold molecular gas, in near isothermal equilibrium
with the cosmic background temperature at $3\,\rm K$ (Pfenniger et
al.~1994). This gas forms a hierarchical structure, possibly a fractal
of dimension $D<2$, spread over more than six orders of magnitude in
scale ($\approx 100 \,\rm pc$ down to $10\,\rm AU$). At the smaller
scale, the tiniest fragments or ``clumpuscules'' are evolving in the
adiabatic regime (Pfenniger \& Combes 1994). They have an average
density of $10^{9-10}\,\rm cm^{-3}$, a column density of
$10^{24-25}\,\rm cm^{-2}$, and a Jupiter mass ($10^{-3}\,\rm
M_\odot$). Their surface filling factor is less than 1\% (simply
estimated by the ratio of the average Galactic and clumpuscule column
densities).  Due to the intrinsic inhomogeneity of the fractal
structure, the collision time-scale of the clumpuscules and larger
clumps is shorter than their collapse time.

So the envisaged equilibrium is dynamical, constantly renewed
collisions in a slight supersonic regime prevent statistically most of
the otherwise expected collapses that would lead to jupiter or star
formation.  Note that the collision speed at the bottom of the
hierarchy is of the order of $0.2\,\rm km\,s^{-1}$, so ``shocks'' do
not imply necessarily strong heating in such conditions.  The frequent
collisions mean also that the average temperature and density of the
clumpuscules are associated with large fluctuations, the warmer tail
of the distribution being just the observed HI.  Indeed collisions not
only tend to heat by mechanical compression, they also simultaneously
{\it cool\/} by gravitational coupling, an often overlooked effect due
to the negative specific heat of gravitating systems.  Of course in
such a model a 100\% inefficiency of jupiter or star formation can not
be excluded.  However we expect that for the same yet poorly
understood reasons that prevent normal molecular clouds to form stars
efficiently
(as a naive expectation about collapsing gas would lead to believe),
similar mechanisms must exist in low excitation regions such as outer
galactic disks and Malin 1 type galaxies preventing the 
the fast cooling HI to collapse immediately.

This model explains naturally the evolution of galaxies along the
Hubble sequence. Through gas accretion, the dark matter in the outer
parts of late-type galaxies is progressively transformed into stars
while galaxies evolve towards earlier types. It also explains the near
constant surface density ratio between the HI gas and dark matter in
the outer parts of spirals (Bosma 1981). During galaxy interactions,
some of the cold gas of the outer parts is progressively heated, and
accounts for the huge amounts of hot gas detected through X-ray
emission in rich galaxy clusters.

With similar argumentation variant models have also been recently
proposed, in which H$_2$ gas contributes significantly to the dark
matter, based mainly in massive proto globular cluster clouds possibly
mixed with brown dwarfs (de Paolis et al.~1995, Gerhard \& Silk 1996).
In this case, the cold H$_2$ gas is not {\it very\/} cold, but has a
temperature between 5 and $20\,\rm K$, which makes it easier to detect
in emission.  However, in the mixed model it is hard to see how any
substantial fractions of cold gas and brown dwarfs can coexist without
virializing rapidly the gas temperature to much higher values than
$5-20\,\rm K$, via dynamical friction and cluster core collapses.

If most of the mass of the Galaxy is under the form of molecular gas,
are there more or less direct ways to detect it?  In this article, we
review several, some novel, possibilities, and try to select the most
promising ones.  To our knowledge a similar earlier attempt was made
only in the 60's (Gould \& Harwit 1963), so the tremendous changes
made in between justify a new article on the subject.  We first
describe the ultrafine structure of the ortho-H$_2$ in its ground
state, due to the interaction between the nuclear spin ($I=1$) and the
magnetic moment induced by the nuclear rotation ($J=1$), which splits
the fundamental in three sublevels ($F=0$, 1 and 2).  Then we consider
trace molecules, formed in the big-bang nucleosynthesis, HD, LiH,
which could be seen in emission due to their electric dipole
rotational transitions, or ions such as H$_2^+$ that could be detected
through its hyperfine structure.  Also, traces of metal enrichment of
this quasi-primordial gas could help to detect it through CO emission
or absorption. Since very cold gas is more likely to be detected in
absorption, we evaluate the probability to detect UV absorption lines
of H$_2$ in front of remote quasars. Finally, we discuss the
submillimeter continuum emission of very cold dust.

The possibility to trace cold gas through their interaction with
cosmic rays and production of gamma rays will be discussed in a
subsequent paper
in which the highly inhomogeneous gas distribution of the ISM will be
taken into account.

\section{ The hyperfine structure of the ortho-H$_2$ }
The hydrogen molecule can be found in two species, the para-H$_2$, in
which the nuclear spins of the two protons are anti-aligned, and the
resulting spin $I=0$. Then, the spin wave function is anti-symmetric,
and the nuclear wave function must be symmetric. Since the two
electrons are paired in the molecule, their total spin and angular
momentum is zero. The angular momentum of the molecule is then only
due to the rotation. In the fundamental state $J=0$, for which there
is a symmetric wave-function.

For the ortho-H$_2$, on the contrary, the total nuclear spin is $I=1$,
since the spins of the two protons are parallel. The spin wave
function being symmetrical, the nuclear radial wave function must be
antisymmetric, by Pauli exclusion principle. Then, the fundamental
state of ortho-H$_2$ is $J=1$.

\subsection{ The energy split }

The rotation of the molecule creates a rotational magnetic moment
parallel and proportional to the angular momentum $J$, since charged
particles (the protons) are in rotation. This rotational moment
interacts with the nuclear spin $I$, through an interaction of the
form $k \vec I\cdot \vec J$. In other words, this comes from the
interaction of $M_I$, the nuclear spin magnetic dipole, with the
magnetic field created from the motion of charges due to rotation. To
this interaction, must be added the spin-spin magnetic interaction for
the two nuclei, and the interaction of any nuclear electrical
quadrupole moment with the variation of the molecular electric field
in the vicinity of the nucleus (Kellog et al.~1939, 1940; Ramsey
1952). In negligible external magnetic field, the H$_2$ states can be
expressed by the reference basis of $F$, $m_F$, with $F=0,1,2$.
Magnetic dipole transitions are possible for $\Delta F =1$, i.e.,
there are two transitions, $F=2-1$, and $F=1-0$. Measurements of these
frequencies in the laboratory have been carried out through magnetic
resonance on molecular beams (Kolsky et al.~1952; Harrick \& Ramsey
1953). The wavelength of these two transitions have been measured at
0.5 and $5.5\,\rm km$, or more precisely at frequencies of
$546.390\,\rm kHz$ and $54.850\,\rm KHz$ respectively for $F=1-0$ and
$F=2-1$.

The reason of such a low energy split of the hyperfine structure is
due to the only interaction between nuclear momenta.  When the fine
structure is due to interaction between electronic momenta, the usual
hyperfine structure corresponds to interaction between electronic and
nuclear momenta. The fine structure splitting is then proportional to
the Bohr magneton $\mu_o = e h/(4\pi m_{\rm e} c)$ squared, where
$m_{\rm e}$ is the electron mass.  The hyperfine structure involves
the product of $\mu_{\rm o}$ with the nuclear magneton $\mu_{\rm n} =
e h/(4\pi m_{\rm p} c)$, where $m_{\rm p}$ is the proton mass. When
the interaction involves two nuclear momenta, the splitting is
proportional to $\mu_{\rm n}^2$, i.e.~smaller than the hyperfine
structure by roughly a factor $m_{\rm p}/m_{\rm e}$.  This could be
called ultrafine structure (cf.~Field et al.~1966).

\subsection{The ortho-para ratio}
Only the ortho-H$_2$ is concerned by the ultrafine structure. Normal
molecular hydrogen gas contains a mixture of the two varieties, with
an ortho-to-para ratio of 3 when the temperature is high with respect
to the energy difference of the two fundamental states ($171\,\rm K$).
For instance, in the warm H$_2$ gas detected by ISO-SWS, an
ortho-to-para ratio of 3 is compatible with observations (Wright et
al.~1996).  At lower temperatures, the ortho-to-para ratio must be
lower, if the thermodynamical equilibrium can be reached, until all
the hydrogen is in para state at $T=0$. However, due to the rarefied
density of the ISM, the ortho-to-para ratio is frozen to the
H$_2$-formation value (Schaefer, private communication).  Considerable
densities are required for the ortho-to-para conversion, which occurs
in solid H$_2$ for instance.

But the fractal gas must be seen as a dynamical structure far out of
thermodynamical equilibrium, not only with large density contrasts,
but also large temperature contrasts, extending to low temperature the
well know inhomogeneous properties of the ISM where it can be well
observed, i.e.~above about $10\,\rm K$.  The HI is then the warm
interface, and H$_2$ the coldest component, in a mass ratio of about 1
to 10.  By continuity, H$_2$ forms from HI and vice versa with a rate
given by the clump collision time-scale at the scale corresponding to
the virial temperature of the transition.  This is in any case
relatively short: at a scale corresponding to a temperature of
$3000\,\rm K$, and a fractal dimension $D\sim1.7$, we estimate a
duty-cycle time of transformation HI to H$_2$ of the order of
$10^{6-7}\,\rm yr$, the cold and warmer phases must be chemically well
mixed.  This is an important difference with the alternative cold gas
models of de Paolis et al.~and Gerhard \& Silk, were gas clumps,
modeled as classical gas blobs, require stability over a substantial
fraction of a Hubble time.

Now the key role in ortho to para conversion in interstellar clouds is
the proton exchange reaction (${\rm H}^+ + {\rm H}_2(j=1) \to {\rm
H}^+ + {\rm H}_2(j=0)$, cf.~Dalgarno et al.~1973; Gerlich 1990).  This
reaction can transform the ortho in a time-scale $5\cdot 10^{13}\,
n({\rm H}_2)^{-1/2}\,\rm s$, if the H$^+$ ions in dense clouds are
essentially due to cosmic ray impacts, with the ionizing flux $\xi =
10^{-17}\,\rm s^{-1}$ characteristic of the solar neighbourhood. The
corresponding time-scale for a clumpuscule near the Sun is $10\,\rm
yr$, and the ortho fraction is negligible, but at large distances in
the Galaxy outskirts, where the cosmic-ray flux is expected to fall
exponentially to zero, we can expect a significant part of ortho-H$_2$
in the cold gas.

\section{ Detectability of the H$_2$ ultrafine lines }
On the Earth, the ionosphere is reflecting the long wavelength
radiations, and this phenomenon is used for radio transmission all
over the world (communications with submarines, for example, can be
made through km wavelengths signals).  The ionospheric plasma is
filtering all frequencies below the plasma frequency $\omega = e (4\pi
n/m_{\rm e})^{1/2} \approx 100\,\rm MHz$.  It is therefore necessary
to observe from space. Even from space, the long wavelength radiations
are somewhat hindered by interplanetary or interstellar scintillations
(e.g.~Cordes et al.~1986). This means, as shown below, 
that spatial resolution above $1^{\circ}$ could not be obtained below
$0.1\,\rm MHz$.

\subsection { Interstellar plasma }
The plasma frequency in the interstellar medium can be estimated by
$\nu_{\rm p} = 9\, n_{\rm e}^{1/2}\,\rm kHz$, where $n_{\rm e}$ is the
electron density in $\rm cm^{-3}$.  Since the latter is in average of
the order of $10^{-3}\,\rm cm^{-3}$, the plasma frequency $\nu_{\rm p}
\approx 250\,\rm Hz$. Radiation of frequencies below that value does
not propagate in the medium. More precisely,
since the ISM is far from homogeneous, low-frequency radiation
propagates in rarefied regions, and is reflected and absorbed by
denser condensations.  For the kilometric wavelengths that we are
interested in, there is no problem of propagation, but the waves are
scattered due to fluctuations in electron density. The electric vector
undergoes phase fluctuations, since the index of refraction is
$(1-\nu_{\rm p}^2/\nu^2)^{1/2}$, where $\nu$ is the radiation
frequency.  If the ISM is modeled by a Gaussian spatial distribution
of turbulent clumps of size $a$, the scattering angle can be expressed
by:
\begin{equation}
\theta_{\rm scat} \approx 
  10^8 \left(\frac{L}{a}\right)^{1/2} 
	\frac{\langle \Delta n_{\rm e}^2\rangle^{1/2}}{\nu^2} 
\rm\  {radian}
\end{equation}
where $L$ is the total path crossed by the radiation, and $\nu$ is the
frequency in Hz (e.g.~Lang 1980). At a typical distance of $L = 3\,\rm
kpc$, and for the frequencies considered ($\approx 200\,\rm kHz$), the
scattering angle is of the order of one degree. The scintillation
problem is therefore severe, and hinders spatial resolution for point
sources, but it is still possible to map the Galactic disk.

The interplanetary medium produces somewhat less scattering, and the 
total order of magnitude remains unchanged. 

\subsection {Intensity of the H$_2$ ultrafine lines}
The radiation has a dipole matrix element proportional to $\mu_{\rm
n}^2$; the line intensity is therefore much weaker than for usual
hyperfine transitions (magnetic dipole in $\mu_{\rm o}^2$). Since the
spontaneous emission coefficient $A$ is proportional to $\nu^3$, the
life-time of a hydrogen molecule in the upper ultrafine states is much
larger than a Hubble time: $A \approx 10^{-32}\,\rm s^{-1}$. It is
then likely that the desexcitation is mostly collisional. Even at the
$3\,\rm K$ temperature, the upper levels are populated in the
statistical weights ratio. A weak radiation is therefore expected, but
the velocity-integrated emission ($\int T_a \,dv$) is ten orders of
magnitude less than for the HI line, for the same column density of
hydrogen.  The prospects to detect the lines are scarce in the near
future, since it would need an instrument of about 6 orders of
magnitude increase in surface with respect to nowadays ground-base
telescopes!  A solution could be to dispose a grid of cables or array
of dipoles spaced by $\lambda/4 \approx 125\,\rm m$ on a significant
surface of the Moon, e.g.~an area of $(300\,\rm km)^2$.  This
requirement could be released, however, if there exists strong
coherent continuum sources at km wavelengths. The H$_2$ ultrafine line
could then be detected much more easily in absorption, with presently
planned km instruments.

\subsection{Galactic background and VLF projects}
Several groups have studied the possibility to observe the sky in the
very low frequency domain ($\approx 0.5-15\,\rm MHz$); the most
important projects being the NASA Low Frequency Space Array (LFSA)
(Weiler et al.~1988, 1994), and ESA Very Low Frequency Array (VLFA,
ESA Report, SCI96-002). These are space projects, since the Earth
ionosphere is opaque below $15\,\rm MHz$.  The present projects now
consider the possibility either of satellites orbiting the Moon, or an
array of dipoles on the Moon itself, to avoid the strong radio
interferences coming from the Earth. On the Moon, the hidden far-side
surface is favoured, to be completely free of the Earth radio
emissions.  However, the lunar ionosphere might add some problems,
especially on the lunar day, so the observations would confined to the
14 days lunar night.

The limit at low frequency ($0.5\,\rm MHz$) is fixed by interstellar
free-free absorption: the optical depth depends on the emission
measure of the medium ($\int N_{\rm e}^2 dx$) and its electron
temperature ($T_{\rm e}$). Typically, the depth $l$ that we can see
through the medium below $\tau =1$ is a function of frequency $\nu$ in
MHz: $l\sim (50\nu)^2\,\rm pc$. At the lower frequency $0.5\,\rm MHz$,
the observations will just cross the Galactic plane, and be able to
see extragalactic sources. One cannot rely on the clumpy structure of
the medium to see through it (as for dust in the optical domain),
since the scattering effects of the turbulent plasma broaden any
emission to an angle $\theta \sim 22' \nu^{-2}$, i.e.~$\sim 1.5^\circ$
at $0.5\,\rm MHz$.

The sensitivity of such arrays depends only on their total surface and
filling factor $f$, but not on the receivers quality, since the system
temperature is dominated by the sky noise itself (galactic background
radiation, essentially of synchrotron origin), which is about $ 3\cdot
10^7\,\rm K$ at $0.5\,\rm MHz$. Arrays of short dipoles ($10\,\rm m$)
on the Moon covering $100\,\rm km$ diameter surface are considered,
with a filling factor of $f \sim 10^{-3}$. The signal to noise ratio
reached in an integration time $t$ in a frequency band $\Delta\nu$ is
then ${\rm SNR} = f \sqrt (\Delta\nu t)$, and can reach ${\rm SNR} =
200$ in $t= 14\,\rm days$.  In other words, we could detect signals at
a few percent of the galactic background within a lunar night.

In front of such an intense Galactic background, the H$_2$ molecules
are absorbing at their line frequency of $0.5\,\rm MHz$. However the
optical thickness could be at most $\tau=10^{-5}$ in condensed
regions, and smoothed to an average of $\tau=10^{-10}$ in regions
subtended by the effective resolution of $\sim 1.5^\circ$.
Observations of an absorbing signal will only be possible towards very
strong continuum sources such as Cas-A, and in any case will be quite
difficult.

\section {The HD and LiH Transitions and Detectability}
HD has a weak electric dipole moment, since the proton in the molecule
is more mobile than the deuteron; the electron then does not follow
exactly the motion of the positive charge, producing a dipole. This
moment has been measured in the ground vibrational state from the
intensity of the pure rotational spectrum by Trefler \& Gush (1968):
its value is $5.85 \pm 0.17\cdot 10^{-4}\,\rm Debye$. The first
rotational level is at $\approx 130\,\rm K$ above the ground level,
the corresponding wavelength is $112\,\mu$. This line could be only
observed in emission from heated regions, and given the very low
abundance ratio HD/H$_2 \approx$ 10$^{-5}$ and weak dipole, does not
appear as a good tracer of the cold gas.

The LiH molecule has a much larger dipole moment, $\mu = 5.9$ Debye
(Lawrence et al.~1963), and the first rotational level is only at
$\approx 21\,\rm K$ above the ground level, the corresponding
wavelength is $0.67\,\rm mm$ (Pearson \& Gordy 1969; Rothstein
1969). The line frequencies in the submillimeter and far-infrared
domain have been recently determined with high precision in the
laboratory (Plummer et al.~1984, Bellini et al.~1994), and the great
astrophysical interest of the LiH molecule has been emphasized
(e.g.~Puy et al.~1993). A tentative has even been carried out to
detect LiH at very high redshifts (de Bernardis et al.~1993).  This
line is unfortunately not accessible from the ground at $z=0$ due to
H$_2$O atmospheric absorption. This has to wait the launching of a
submillimeter satellite, but is a good candidate. The abundance of LiH
has been recently estimated to a very low value ($< 10^{-15}$) in the
post-recombination epoch (Stancil et al.~1996). The rate coefficient
for LiH formation through radiative association is now estimated 3
orders of magnitude smaller than previously (Lepp \& Schull 1984).  In
very dense clouds, however, three-body associations reactions have to
be taken into account, and most of the lithium turns into molecules;
an order of magnitude for LiH abundance is then LiH/H$_2 \approx
10^{-10}$, and the optical depth should reach 1 for a column density
of $10^{12}\,\rm cm^{-2}$, or N(H$_2) = 10^{22}\,\rm cm^{-2}$, in
channels of $1\,\rm km\,s^{-1}$. The line should then be easily
detectable in normal molecular clouds, within the optical disk of the
Galaxy; the more so as the primordial abundance of Li could be
increased by about a factor 10 in stellar nucleosynthesis (e.g.~Reeves
1994).  In the outer Galaxy parts of course, the same problems of
excitation arise for a very cold gas, and the surface filling factor
might be a problem for absorption measurements.

\section {The H$_2^+$ hyperfine transitions}
The abundance of the H$_2^+$ ion is predicted to be no more than
$10^{-11}$ to $10^{-10}$ in chemical models (e.g.~Viala 1986). But the
H$_2^+$ ion possesses an hyperfine structure in its ground state,
unfortunately in the first rotational level $N=1$. The electron spin
is 1/2, and the nuclear spin $I=1$, which couple in $F_2 = I + S= 1/2$
and $3/2$; then $F= F_2 + N = 1/2$, $3/2$ and $5/2$. Five transitions
are therefore expected, of which the strongest is $F$, $F_2= 5/2$,
$3/2 \to 3/2$, $1/2$, at $1343\,\rm MHz$ (Sommerville 1965; Field et
al.~1966). At the interface between the cold molecular gas and the
interstellar/intergalactic radiation field, one can hope to encounter
a sufficient column density of H$_2^+$. The excitation to the $E_u =
110\,\rm K$ level is problematic however.

\section{C and O enrichment of the quasi-primordial gas}
\subsection{CO in emission or absorption}
The usual tracer of molecular hydrogen in the ISM is the CO molecule,
but is valid only for enriched gas, and fails at large distances from
galaxy centers. The abundance [O/H] decreases exponentially with
radius in spiral galaxies, with a gradient between $-0.05$ and
$-0.1\,\rm dex/kpc$ (e.g.~Pagel \& Edmunds 1981), and the
N(H$_2$)/I(CO) conversion ratio is consequently increasing
exponentially with radius (Sakamoto 1996). There could be even more
dramatic effects such as a sharp threshold in extinction (at 0.25 mag)
before CO is detectable (Blitz et al.~1990), due to
photo-dissociation. The effect of metallicity on the conversion ratio
has long been debated (e.g.~Elmegreen 1989), but strong evidence
exists in the Magellanic Clouds (Rubio et al.~1993), and in nearby
dwarf galaxies (Israel, Tacconi \& Baas 1995; Verter \& Hodge 1995;
Wilson 1995).  The fact that the CO line is optically thick does not
make it insensitive to metallicity, since the observed size of
molecular clouds is related to the $\tau=1$ surface, and clouds appear
much smaller at low metallicity. This fact is often ignored in
theoretical models, where clouds or clumps are assumed to have a
constant column density with radius (e.g.~Wolfire et al.~1993).

It is easy to estimate until which radius the dense clouds are likely
to contain CO molecules, if we estimate that the opacity gradient
follows the metallicity gradient. Assuming the proportionality
relation N(H)$\approx 2\cdot 10^{21}\,\rm A_v\, atoms\, cm^{-2}\,
mag^{-1}$ between the gas column density and opacity in the solar
neighbourhood (Savage et al.~1977), and a column density of
$10^{25}\,\rm cm^{-2}$ for the densest fragments, their opacity
A$_{\rm v}$ falls to 0.25 at $R \approx 60\,\rm kpc$, but of course
the CO disappears on large scales before.

Another effect hinders the detection of molecular tracers in emission,
far from star-formation regions, which is the lack of heating
sources. It is impossible to detect emission from a cloud at a
temperature close to the background temperature. Only absorption is
possible, although difficult for a surface filling factor of less than
1\%. Absorption is biased towards diffuse clouds, or intercloud
medium, which has a large filling factor, and a low density (and
therefore a low excitation temperature).  This is beautifully
demonstrated in the molecular absorption survey of Lucas \& Liszt
(1994, 1996) in our Galaxy: practically all of the absorbing clouds
are diffuse (A$_v \la 1$) with a low excitation temperature. When the
gas becomes metal deficient, the diffuse medium is preferentially
depleted in molecules, through UV photo-dissociation.

\subsection{Intracluster gas}
It might still be possible in special environments such as galaxy
clusters, to try to detect the cold gas, polluted by the enriched
intra-cluster medium.  The problem is to estimate the actual
enrichment, based on the gas mixing.  The medium is expected to be
multi-phase from theoretical calculations (e.g.~Ferland et al.~1994),
and direct EUVE observations of the $5\cdot 10^5\,\rm K$ gas in the
Virgo cluster (Lieu et al. 1996) support a substantial cooling of the
X-ray gas toward cold phases.  In the outer parts of the cluster,
before galaxy interactions have heated the cold gas around individual
galaxies, the latter is still cold and metal deficient.  Progressively
the cold gas is heated and transfered in the hot X-ray emitting phase,
where it is metal enriched, at least to the intra-cluster abundance of
$\approx 0.3\,\rm Z_\odot$. In the cluster center, where the density
of hot gas is high enough, the hot phase becomes unstable to cooling,
and generates the cooling flow.  Since there is no reason why the
ubiquitous hierarchical fragmentation observed in the Galaxy is not
universal, we must expect that in galaxy cluster gas fragmentation
occurs too down to low temperature, small sizes and high densities,
i.e.~to molecular clumpuscules (Pfenniger \& Combes 1994). This
accounts for the apparent complete disappearance of gas in cooling
flows, and may explain the high concentration of dark matter in
clusters deduced from X-ray data and gravitational arcs (Durret et
al.~1994; Wu \& Hammer 1993).

Many authors have tried to detect this gas in emission or absorption,
either in HI (Burns et al.~1981; Valentijn \& Giovanelli 1982; Shostak
et al.~1983; McNamara et al.~1990; Dwarakanath et al.~1995) or in the
CO molecule (Grabelsky \& Ulmer 1990; Mc Namara \& Jaffe 1994;
Antonucci \& Barvainis 1994; Braine \& Dupraz 1994; O'Dea et
al.~1994).  Maybe the best evidence of the presence of the cooling gas
is the extended soft X-ray absorption (White et al.~1991).  Although
the HI is not detected in emission with upper limits of the order of
$10^9-10^{10}\,\rm M_\odot$, it is sometimes detected in absorption,
when there is a strong continuum source in the central galaxy. The
corresponding column densities are $>10^{20}\,\rm cm^{-2}$.  There is
also some CO emission in Perseus A (Lazareff et al.~1989), but which
may come from the galaxy itself and not the cooling flow.

To produce the X-ray absorption observed, the gas must have a high
surface filling factor ($\approx 1$), and a column density of the
order of N$_{\rm H} \approx 10^{21}\,\rm cm^{-2}$ (White et
al.~1991). The total mass derived is of the order of $10^{11}\,\rm
M_\odot$ over a $100\,\rm kpc$ region.  This gas can only correspond
to the atomic gas envelope of the denser, low filling factor,
molecular clouds.  If the molecular clouds are cold ($T\approx 3\,\rm
K$) and condensed (filling factor $< 1\%$), it is extremely difficult
to detect them, either in emission or in absorption, even at solar
metallicity. The best upper limits reported in the literature ($N({\rm
H}_2)< 10^{20}\,\rm cm^{-2}$, average over regions $10\,\rm kpc$ in
size, but assuming $T\approx 20\,\rm K$ and solar metallicity,
i.e.~the standard $N({\rm H}_2)/I$(CO) conversion ratio, are perfectly
compatible with the existence of a huge cold H$_2$ mass (the
conversion factor tends to infinity when the temperature tends to the
background temperature).  The HI gas, on the contrary, could give much
significant constraints, since it is a warm gas with high filling
factor. The column density detected in absorption are all compatible
to what is expected from the cooling flows. The HI gas is observed
redshifted, falling on the central galaxy. Why is there no emission?
It could be that the absorbing gas is too cold, and/or part of the gas
absorbing X-rays is ionized (as observed H$\alpha$ filaments suggest).

That the molecular gas can cool down to $\approx 2.7\,\rm K$ within
the intra-cluster environment has been demonstrated by Ferland et
al.~(1994), but debated (e.g.~O'Dea et al.~1994).  O'Dea et al.~(1994)
evaluate the X-ray heating rate of molecular clouds, assuming that the
attenuating column density can be no more than $10^{21}\,\rm cm^{-2}$
($\tau=0.2$ to 2) measured by White et al.~(1991). But this assumes
that the absorbing medium is homogeneous.  On the contrary, if a
hierarchical structure of dense molecular clouds is embedded within
this HI envelope, the attenuating column density can be 3 or 4 orders
of magnitude ($\tau \gg 1$) locally ($f<1\%$); the molecular fragments
can be completely screened from the X-ray flux. The presence of the
X-ray flux introduces a warm ionized interface between the very cold
molecular gas and the hot medium. That might be this phase that is
seen in soft X-ray absorption by White et al.~(1991). The HI
absorption measurements reveal that only a small part of this
interface is neutral.

Recently, Lieu et al (1996) and Bowyer et al (1966) have detected large
quantities of gas at intermediate temperature of $5\cdot 10^5\,\rm K$
in the Virgo and Coma clusters with the EUVE satellite (Extreme
UltraViolet Explorer).
Since this gas is cooling very rapidly, being near the peak of the
radiative cooling curve, it should be very transient. Although its
extension coincides with that of the cooling flow, the mass flow
involved would be 30 times that of the cooling flow itself. Its
detection is therefore not only evidence for a continuous flow of gas
cooling from the high virial temperature of the clusters (10$^7$ K),
but also of possible other heating mechanisms like shocks.  Also the
detection of the near-infrared quadrupolar emission line
H$_2$(1-0)S(1) in central cluster galaxies with cooling flows (and
their non-detection in similar control galaxies without cooling flows)
strongly confirms the gas flow as it passes through the temperature of
2000K (Jaffe \& Bremer 1997).

\section {H$_2$ absorption lines in front of quasars}
Absorption should be the best way to trace the cold gas. If the
effective gas temperature is indeed close to the background $T_{\rm
bg}$, emission is extremely weak. The antenna temperature is, in the
Rayleigh-Jeans approximation:
\begin{equation} 
T_{\rm a} = 
  \left( T_{\rm ex}-T_{\rm bg} \right) \left(1 - {\rm e}^{-\tau}\right)
\end{equation}
where $T_{\rm ex}$ is the gas excitation temperature and $\tau$ the
optical thickness in the transition considered. On the contrary,
absorption is biased towards cold gas, since the observed signal is
proportional to $N_0/T_{\rm ex}$, where $N_0$ is the column density in
the ground state.

\subsection {Detected H$_2$ absorption lines}
Absorption in the vibration-rotation part of the spectrum (in
infrared) is not the best method, since the transitions are
quadrupolar and very weak. An H$_2$ absorption in Orion has been
detected only recently (Lacy et al.~1994) and the apparent optical
depth is only about 1\%. This needs exceptionally strong continuum
sources, which are rare.  Electronic lines in the UV should be more
easy to see in absorption.

However, recognizing molecular hydrogen absorption in UV spectra of
quasars is not easy due to the dense Ly$\alpha$ forest.  Many
tentatives have remained inconclusive.  A careful cross-correlation
analysis of the spectrum is needed in order to extract the H$_2$ lines
from the confusion.  Already Levshakov \& Varshalovich (1985) had made
a tentative detection, towards PKS$\,0528-250$, with some 13
coincident lines among the Lyman and Werner H$_2$ bands.  But
molecular hydrogen has been firmly found in absorption in front of the
quasars PKS$\,0528-250$ (Foltz et al.~1988) and QSO$\,0013-004$ (Ge \&
Bechtold 1997).  These objects are at redshifts $z=2.8$ and $z=1.97$
respectively and are viewed through damped Ly$\alpha$ systems; the
inferred H$_2$ column densities are $10^{18}\,\rm cm^{-2}$ and $7\cdot
10^{19}\,\rm cm^{-2}$, with estimated widths of 5 and $15\,\rm
km\,s^{-1}$ and kinetic temperatures of 100 and $70\,\rm K$. These
conditions, together with the molecular fraction derived $f_{\rm H_2}=
0.002$ and $0.22$, indicate a diffuse medium.

It should be remarked that only the very low column densities of H$_2$
can be detected this way, since for moderate to high molecular column
densities, the line damping is so high that the quasar is no longer
visible, as is developed in next paragraph. This happens in only rare
cases, since the filling factor of high H$_2$ surface density gas is
lower than $f=1\%$. Unfortunately, it is not presently possible to
observe H$_2$ absorptions at $z=0$ from our own Galaxy in front of
quasars, since the HST has no instrument at the right frequency
(wavelengths smaller or equal to $1100\, \AA$). The probability to
have an intervening high molecular column density is of course very
small at non-zero redshift.  Moreover, for redshift high enough, the
size of a dense clump is not large enough to cover the quasar UV
continuum, and the depth of the absorption is not larger than
$f=1\%$. The possibility to detect H$_2$ absorption from our own
Galaxy has to await future UV satellites, such as the Lyman-FUSE
(Far-Ultraviolet Spectrographic Explorer) mission.

\subsection{Line damping }
For a high column density of H$_2$, such as $10^{25}\,\rm cm^{-2}$ as
should be the case at the center of a clumpuscule, the absorption
lines in the UV range is extremely saturated.  In these circumstances,
the actual width of the lines exceeds considerably the Doppler width,
to be entirely dominated by the natural width.  Radiation damping
determines the profile. While the Doppler width should be of the order
of $0.1\,\rm km\,s^{-1}$, the equivalent width must be computed from
the ``square-root'' region of the curve of growth. The expected
equivalent width $W$ normalised to the wavelength $\lambda$ can be
crudely estimated to be $W/\lambda \sim 1$. This means that all H$_2$
lines in the Lyman and Werner bands are overlapping, and all the UV
and optical light are absorbed. A special computation of the line
profile should then been done, since the simple Lorentzian profile is
no-longer valid far away from the resonance line (when $\Delta\nu \sim
\nu$).

We have derived the total cross section of the interaction
photon-molecule, including absorption and scattering.  It can be shown
(see Appendix) that this total cross-section, from the fundamental
state, and in the vicinity of electronic dipolar transitions, can be
written as:
\begin{equation}
\sigma(\omega) = { {\omega}\over {\epsilon_0 c \hbar} } 
	\sum_i {|d_{0i}|^2} 
	\frac{ \left(\frac{\omega}{\omega_{0i}}\right)^3 \frac{\Gamma_i}{2} }
     		{(\omega-\omega_{0i})^2 +
\left(\frac{\Gamma_i}{2}\right)^2}  
\end{equation}
where the resonances occur for $\omega=\omega_{0i}$ (with
corresponding dipole matrix element $d_{0i}$), and $\Gamma_i$ is the
natural width of the $i$-level. At low frequencies, the classical
Rayleigh scattering formula in $\omega^4$ is retrieved.

\begin{figure}
\hskip-7mm
\psfig{angle=-90,figure=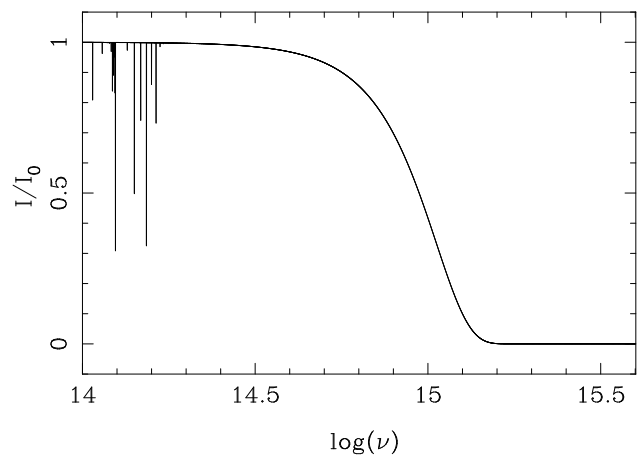,width=7.5cm}
\vskip-3truecm
\caption[]{Simulation of the absorption profile expected from a
clumpuscule in the Milky Way in front of a remote UV quasar.  Plotted
is the ratio $I/I_0$ of the depth of the absorption normalised to the
continuum. All the H$_2$ UV lines and NIR lines have been taken into
account. The continuum has completely disappeared at high frequencies
($\lambda$ smaller than $3000\,\AA$, for a non-redshifted quasar).}
\label{damping}
\end{figure}

This cross section has been taken into account to compute the expected
profile in front of a clumpuscule (Fig.~\ref{damping}).  At high
frequency, the asymptotic limit of the total cross-section is the
Thomson cross-section $\sigma_T = 6.6\cdot 10^{-25}\,\rm cm^{-2}$.
The high frequency wing of the H$_2$ lines is then always optically
thick.  The clumpuscule is absorbing all light with wavelength smaller
than the optical band.

Within this band, the H$_2$ clumpuscules play then the role of dusty
globules, removing the whole continuum of the region (even the He
lines could not be seen). Since their surface coverage is not large,
however, they do not completely obscure the quasar source, depending
on its size, but only a few percents of it. The type of absorption
depends strongly on the relative distances.  If the absorbing clouds
belong to an intervening galaxy at high redshift ($z\ge 0.5$), the
angular size of the obscuring clumpuscules is less than a
micro-second, and the effect on the remote quasar would remain
un-noticed.  If the absorption comes from our own Galaxy, the angular
size can be a fraction of an arcsecond, and the optical QSO could be
entirely covered and obscured. Through proper motions of the absorbing
globules, at about $100\,\rm km\,s^{-1}$, this absorption can appear
or disappear on scales of one year.

\subsection{ Helium lines }
The most abundant primordial element after hydrogen is helium, which
remains atomic in dense molecular clouds. We have not considered it as
a good tracer candidate: only the UV lines could be detected in
absorption, the first one being at $584\,\AA$. The intensities of the
lines are similar to that of molecular hydrogen, the problems of line
damping and of filling factors are comparable. The direct H$_2$ lines
are therefore a better candidate. When the latter are damped, the He
lines are also obscured. Only in the case of complete condensation of
the H$_2$ molecules into
ice, does the He lines become useful.  Helium is then the most
abundant gas, its pressure supporting the clumpuscule against a
free-fall gravitational collapse.

\section{Submillimeter Continuum }
The far-infrared and submillimeter continuum spectrum from $100\,\mu$
to $2\,\rm mm$ has been derived from COBE/FIRAS observations by Wright
et al.~(1991) and Reach et al.~(1995).  They show that in addition to
the predominant warm dust emission, fitted by a temperature ranging
from $T\approx 16$ to $21\,\rm K$, according to longitude, there is
evidence for a very cold component, with temperatures between $T = 4$
and $7\,\rm K$, ubiquitous in the Galaxy, and somewhat spatially
correlated with the warm component. The opacity of the cold component,
if interpreted by the same dust emissivity model (varying with
frequency as $\nu^2$, is about 7 times that of the warm component.
Most of the gas mass would then be contained in this component, which
is, relatively to the warm component, stronger at high latitude and
away from the galactic center. This is clearly seen in the longitude
dependence of the cold/warm power ratio, which shows a characteristic
minimum towards zero longitude (Reach et al.~1995).

\subsection{H$_2$ dimers}
Schaefer (1996) proposes that the cold dust component detected by
COBE/FIRAS might be due in fact to molecular hydrogen emission, as
collision-induced dipole transitions: in small very high density
fluctuations (at a fraction of ``amagat'', i.e.~a density of $4\cdot
10^{18}\,\rm cm^{-3}$), the H$_2$ gas can emit a continuum radiation,
corresponding to free-bound or free-free transitions of weakly and
temporarily bound H$_2$ molecules, the dimers, and containing a large
fraction of ortho-H$_2$.  By symmetry the para-para complexes do not
produce such a radiation.  A good fit is found for the COBE spectra if
the dense H$_2$ clouds follow the HI distribution in the outer parts
of the Galaxy.  At least this shows that a cold component (``dust'')
is indeed associated to the warmer HI.

This weak-dipole radiation due to H$_2$ collisional complexes is an
interesting possibility to detect the presence of cold molecular
hydrogen. This kind of radiation has been identified in planetary
atmospheres during the Voyager IRIS mission (Hanel et al.~1979).  Only
exceptionally dense regions could explain the signal detected by COBE,
since the emission is proportional to the square of the density
(Schaefer 1994). The required density then imposes the temperature
($T>11\,\rm K$), to avoid the transition to solid molecular hydrogen.
Already we had remarked that at the present cosmic background
temperature of $T_{\rm bg0}=2.726\pm 0.01\,\rm K$, the average
pressure in the H$_2$ clumpuscules was about 100 times the pressure of
saturated vapour, and that probably a small fraction of the molecular
mass might be in solid form (Pfenniger \& Combes 1994, and several
references therein about previous papers discussing the possibility of
hiding molecular hydrogen in solid form).  Snow flakes of H$_2$ can
improve the coupling with the CBR, but the large latent heat of
$110\,\rm K$ per H$_2$ molecule and the lack of nucleation sites in
metal poor gas may prevent a large mass fraction to freeze out.  The
condition of dimerization is then largely satisfied in the physical
conditions of the clumpuscules ($T\approx 3\,\rm K$, $n\approx
10^{10}\,\rm cm^{-3}$), and we expect continuum radiation to be
emitted and absorbed by the H$_2$ collisional complexes, through
collision-induced dipole moment.  The absorption coefficient peaks in
the submillimeter domain ($\lambda =0.5\,\rm mm$). The optical depth
of each clumpuscule is however quite low, $\tau \approx 10^{-9}$. This
radiation is not detectable, but for the densest fluctuations, where
the density exceeds $1 \, {\rm amagat} = 4\cdot 10^{18}\,\rm
cm^{-3}$. The total mass required in these fluctuations is already a
few $10^7\,\rm M_\odot$ for the whole Galaxy.  The rather steep
dependence on density of the H$_2$ emission coefficient prevents the
obtaining of a less crude model, the more so as the optical depth of
the dust components are not yet exactly known.

\subsection{Cosmic infrared background}
Reach et al.~(1995) have neglected any cosmic infrared background
(CIBR) in their modeling, but it has been claimed by Franceschini et
al.~(1994) that 5 to 25\% of the sky brightness in the submillimeter
and far-infrared bands is in fact the CIBR coming from the integrated
light of galaxies.  In their model, early-type galaxies at high
redshift are highly obscured, and most of their light coming from the
intense star-formation phase is re-radiated in the FIR. The COBE data
allows a significant fraction of the sky brightness for the CIBR,
since there is some freedom in the relative contributions of the
Galaxy and the isotropic backgrounds.  In any case, however, the
hypothesis of the existence of a CIBR does not solve the cold
component problem. Reach et al.~(1995) have tried, prior to their
analysis, to subtract a CIBR component as high as that computed by
Franceschini et al.~(1994), but this did not change their conclusions
about the existence of the two components, which temperatures were
unaffected.

The basic problem in deriving the exact amount of the CIBR
contribution is the subtraction of the foreground emission.  Another
recent interpretation of the COBE/FIRAS data has been put forward by
Puget et al.~(1996) and Boulanger et al.~(1996). Their approach
consists in modeling the foreground dust emission according to the
Dwingeloo recent HI survey (cf.~Hartmann 1994). They use the fact that
the relation between the integrated HI emission and the far-infrared
emission is almost linear, at least for weak HI emission,
corresponding to column densities $N($HI$) < 4\cdot 10^{20}\,\rm
cm^{-2}$. They consider only the high latitudes: $|b| > 30^\circ$,
where the temperature of the dust can be fitted with a constant
temperature $T=17.5\,\rm K$, with an emissivity varying as
$\nu^2$. After subtracting this component, with the proportionality
factor found between HI and FIRAS emission they also subtract the
infrared emission associated with the ionized gas, equivalent to an
average column density of $0.4\cdot 10^{20}\,\rm H\, cm^{-2}$,
(following a cosecant law variation with latitude). This is necessary,
since there is a bias in the HI-FIR emission relation, i.e., some
far-infrared emission exists at zero HI emission, and this is
interpreted as coming from dust associated to ionized gas, not
correlated with HI. The residual has then the spectrum of a very cold
dust component, peaking around $250\,\mu$.  However this treatment
neglects completely the molecular contribution, which in our Galaxy is
of the same order of magnitude in mass as the HI component. The H$_2$
component distribution cannot be accurately known, since the CO
molecule is not a good tracer.  It was claimed by Puget et al.~(1996)
that they avoid molecular hydrogen by selecting only low column
density of HI ($< 4.5\cdot 10^{20}\,\rm cm^{-2}$).  However they have
first smoothed out the HI map at the FIRAS resolution of
$7^\circ$. The molecular component being very clumpy, its signature is
completely washed out in the process. A typical high latitude cloud
(HLC) detected in molecular transitions is in average of 1 deg$^2$
extent (Magnani et al.~1996), so that its average HI column density of
$10^{21}\,\rm cm^{-2}$ becomes negligible, diluted by a factor 49. The
average surface density of HLC's has been estimated by Magnani et
al.~(1996) already below the HI threshold of $4.5\cdot 10^{20}\,\rm
cm^{-2}$.  This means that it is impossible to avoid molecular
hydrogen at the FIRAS resolution.

\begin{figure}
\vskip-1.5cm 
\hskip-9mm \psfig{angle=0,figure=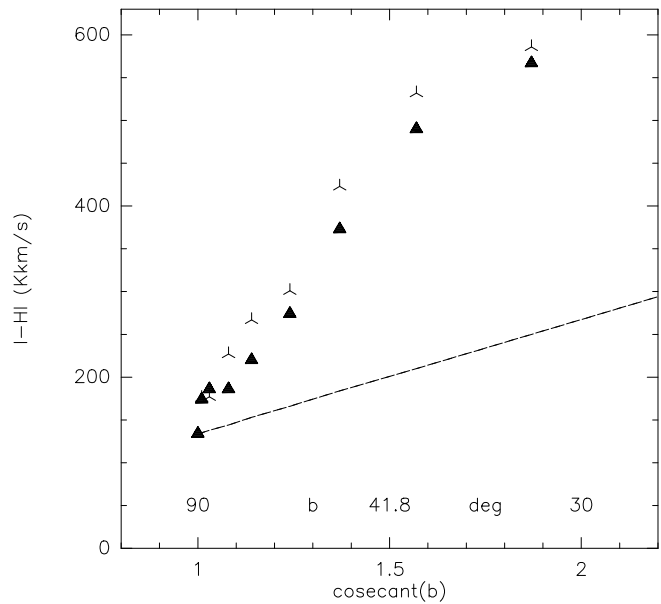,width=10.7cm}
\vskip-1.2cm 
\caption[]{Distribution of the HI emission with latitude in the
Galaxy. The data come from the Bell-Labs survey (Stark et al.~1992).
The crosses correspond to negative latitudes, and the filled triangles
to positive $b$. The dashed line indicates the expected cosecant law,
if the local HI plane could be approximated by a plane-parallel 
component.}
\label{csb}
\end{figure}

\subsection{Uncertainties in deriving the CIBR level}
It is quite difficult to subtract local material emission from a
possible isotropic component, since the distribution of the local
material departs significantly from a plane-parallel geometry. It has
been known from the first HI surveys that the Sun lies in a hole of HI
deficiency (e.g.~Burton 1976). The total column density of HI observed
at the pole is $2\cdot 10^{20}\,\rm cm^{-2}$, while it should have
been twice higher through extrapolation of the $|b|=30-40^\circ$ data.
It is well known from many gas tracers that the Sun appears not too
far from the center of an elongated Local Bubble mostly filled with
hot gas, deficient in neutral gas, and of radius about $60-100\,\rm
pc$ (Welsh et al.~1994), approximately bounding the local ISM (LISM)
(Cox \& Reynolds 1987).  The high latitude clouds observed in
molecular transitions might be at the boundary of this cavity; their
distribution does not show any longitude dependency, while it is also
quite deficient at high latitude (e.g.~Magnani et al.~1996). The
latitude distributions of the HI and IRAS $100\,\mu$ emission
component depart from the expected cosecant law (e.g.~Boulanger \&
Perault 1988, and Fig.~\ref{csb}), which might be attributed to the
continuation of the Local Bubble in a high latitude chimney.

Note that between $|b| = 30$ to $90^\circ$, the region sampled is only
the close neighbourhood of the Sun, within $500\,\rm pc$ or $1\,\rm
kpc$, according to the scale height of the medium (the HI scale height
is $180\,\rm pc$, and the H$_2$ $60\,\rm pc$). Therefore, no longitude
dependence is expected for $|b| > 15^\circ$.

If the neutral warm gas cannot be considered to follow the plane
parallel geometry locally, i.e.~there is a deficiency of warm dust
associated to the HI gas at high latitude, this deficiency could be
compensated by a colder medium, emitting only at the lower FIRAS
frequencies. Indeed, the long wavelength ($200-500\,\mu$) emission
from COBE/FIRAS does not suffer from this depletion, and follows
better the cosecant law.

The correlation between the HI flux and the COBE/FIRAS emissions at
wavelengths between $100\,\mu$ and $1\,\rm mm$ is in fact non-linear,
as found by Boulanger et al.~(1996).  Also there appears to be some
FIR-submm emission even at zero HI flux, which could be due to other
gas components. It is therefore quite difficult to subtract the
emission associated to the HI, since the results depend on the linear
fit used, i.e.~on the maximum $W$(HI) chosen. Boulanger et al.~(1996)
choose a rather low HI column density ($W(\rm HI) = 250 \,K
\,km\,s^{-1}$), which induces strong selection effects as a function
of latitude; indeed when averaged over longitude, the threshold
$W=250\,\rm K\,km\,s^{-1}$ (or $N({\rm HI}) = 4.5\cdot 10^{20}\,\rm
cm^{-2}$) corresponds to $|b|> 42^\circ$. This range of latitude is
not sufficient to reveal a cosecant law, since low column density
regions at lower latitudes is not completely sampled (only the very
low column densities are selected at lower latitudes).

Moreover, the mere subtraction of a linear FIR/HI fit, which
underestimates the FIR at high latitude (low column density) and
overestimates at low latitude (high column density) is able to create
an isotropic residual.

The residual found by Puget et al.~is comparable to the FIR flux at
zero HI obtained through linear fitting of the FIR/HI correlations by
Boulanger et al.~(1996). This residual corresponds to FIR emission not
associated to HI, and could come from colder gas than the HI (for
example molecular). That would explain why its spectrum is shifted to
long wavelengths. At least part of this residual could come from cold
H$_2$ at high latitudes, that has not been properly sampled by CO.

\begin{figure}
\vskip-5truemm
\hskip-6mm
\psfig{figure=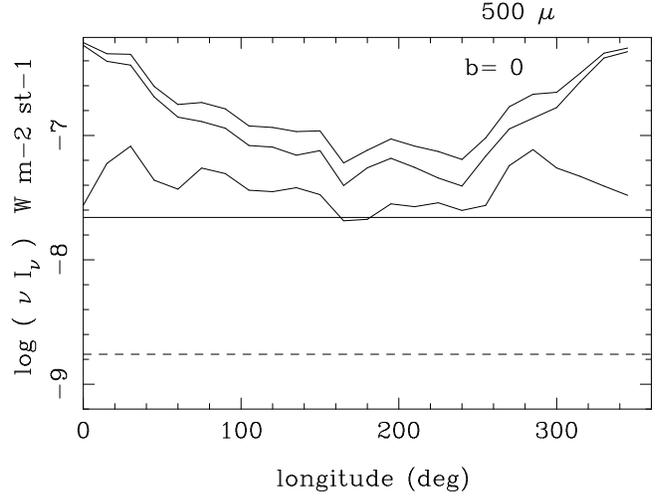,angle=-90,width=7.7cm}
\vskip-28truemm
\caption[]{Relative contribution at $\lambda = 500\,\mu$ of the
various components: the three broken lines correspond to the cold
component (bottom), warm component (middle) and both together (top)
analysed by Reach et al.~(1995).  The dashed horizontal line is the
infrared isotropic background tentatively identified by Puget et
al.~(1996), and the full horizontal line is the CMBR component. At
lower frequencies, $\lambda$ between 500 and $1000\,\mu$ both the cold
component and the CMBR increase relatively to the others. }
\label{reach500}
\end{figure}

\begin{figure}
\vskip-5truemm
\hskip-6mm
\psfig{figure=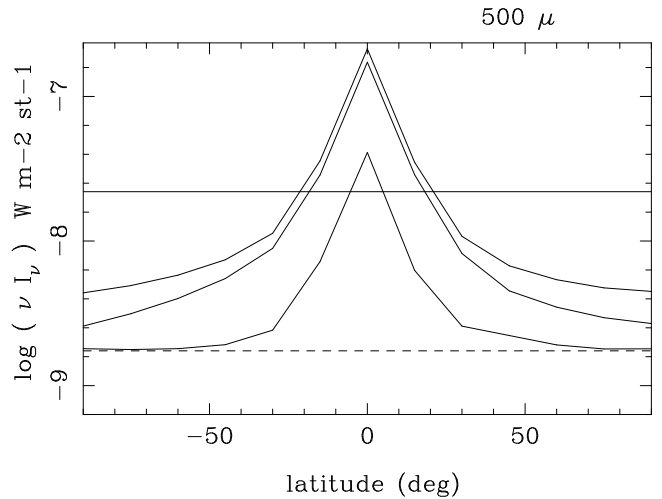,angle=-90,width=7.7cm}
\vskip-28truemm
\caption[]{Same as Fig.~\ref{reach500}, but for the latitude
dependence.  The cold component appears relatively isotropic at $|b|>
50^\circ$. Note that the coincidence with the proposed level for the
CIBR occurs only at $500\,\mu$, since the spectra of the two
components are different.}
\label{reachb}
\end{figure}

The relative contributions of all these components are summarised in
Fig.~\ref{reach500}, where we have plotted as a function of longitude
the flux of the cold and warm component at $b=0^\circ$, as derived by
Reach et al.~(1995).  When the fluxes of the various components are
plotted as a function of latitude (Fig.~\ref{reachb}), it can be noted
that the cold component contribution appears indeed almost isotropic
at $|b|> 50^\circ$.  The cold component is also relatively more
important outside of the Galactic center.  Its longitude distribution
corresponds to what is expected from a ring-like distribution, with a
hole in the Galactic center.

It is important to note (cf.~Fig.~\ref{reach500}) that at $b=0^\circ$,
the cold component found by Reach et al.~(1995) is 50 times higher
than the CIBR, tentatively detected by Puget et al.~(1996). This is
obtained at $\lambda = 500\,\mu$, but the ratio is even higher at
lower frequencies (up to 100 at $1000\,\mu$), since the cold component
has a lower temperature. The existence of the very cold component is
inescapable. It cannot disappear, even in varying extensively the
power $\alpha$ of the emissivity law.

\section{Cold dust associated to dense cores in Giant Molecular Clouds}
One of the best interpretation of the COBE-cold component could be
dust associated to cold molecular hydrogen deep inside Giant Molecular
Clouds (GMC), shielded from the ISRF by the high opacity of the cloud
envelopes. That such cold component should exist is no surprise. It
was already predicted by Mathis et al.~(1983) in their investigation
of the dust emission spectrum in both the diffuse ISM and in GMC's.
From the determination of the dust absorption cross-section from the
Lyman continuum to the submm range, and a model of the ISRF, they
determine the radiation field inside a GMC, as a function of the
optical depth from the surface. Assuming a typical homogeneous cloud,
they find that the ISRF is absorbed in a very thin outer layer of the
cloud. Inside the cloud the ISRF is converted entirely into FIR
radiation, and its mean intensity is about five times the Galactic FIR
emission. Only the thin outer layer of the clouds is heated enough to
radiates at $60-100\,\mu$: it corresponds to the warm dust traced by
IRAS observations. It is well known that this dust mass represents
only about 10\% of the total dust mass (e.g.~Thronson \& Telesco 1986,
Devereux \& Young 1990).  Well inside the GMC's, both silicate and
graphite grains reach a temperature between 5 to $7\,\rm K$ (Mathis et
al.~1983).  This estimation did not include the clumping of gas, which
could be responsible to even larger opacities, and lower grain
temperatures.

If the existence of this cold gas could have been easily foreseen, the
corresponding amount of gas was completely unknown. This is due to the
fact that the molecular transitions are no longer a good tracer of
H$_2$ mass at high density and low temperatures, as shown by studies
of nearby molecular clouds, such as Orion.  Of course, the main
isotopic lines such as CO and CS are highly optically thick in such
conditions, but also the rare isotopes, such as C$^{18}$O or
C$^{34}$S, are not good tracers of molecular condensations, certainly
because of depletion of molecules on grains at high density (chemistry
is not responsible, since the CO molecule is not destroyed at high
density).
 
Mezger et al.~(1992) and Launhardt et al.~(1996) have mapped in the
millimetric dust emission dense cores in the Orion molecular cloud.
They discover a certain number of condensations, where the dust
emission peaks, which correspond to minima in dust temperature.  The
dust condensations {\it are not} or only barely visible in isotopic CO
and CS transitions, probably because molecules have frozen out on dust
grains (Mezger et al.~1992): it appears that molecule depletion
becomes important at densities $n_{\rm H} > 10^6\,\rm cm^{-3}$, at
least at low dust temperatures $T_{\rm d} < 20\,\rm K$.  If the
$1.3\,\rm mm$ emission of dust is a good pointer towards the dense
cold condensations of the Orion GMC, they might not indicate the
actual H$_2$ mass, because of large uncertainties in the physics of
grains and their true opacities, and also in the actual temperature
(they are determined around $15\,\rm K$ from the emission ratio
between 870 and $1300\,\mu$).  Virial masses determined from the CS
line-widths are found $20-30$ times higher than the gas mass derived
from the $1.3\,\rm mm$ dust emission (Launhardt et al.~1996).

The mass in dense cold condensations can therefore only be traced by
dust emission at long wavelength, such as with the COBE/FIRAS
instrument (from $4.5\,\rm mm$ to $104\,\mu$).  Although Reach et
al.~(1995) have discovered precisely such a cold component ($4-7\,\rm
K$), they do not attribute it to cold dust, because it is also present
at high latitude. Their argument is that these cold condensations must
then be seen in the extinction maps at high latitude, and also their
surface should be $100\,\mu$ emitters in IRAS maps, and they are not
observed. But these clouds are actually detected, both in the IRAS
maps, and in extinction maps, and correspond to the well-known high
latitude molecular clouds (HLC, Magnani et al.~1985). The surface
filling factor of these HLC's has been debated.  Magnani et al.~(1986)
estimated a surface filling factor of $5\cdot 10^{-3}$, while
Heithausen et al.~(1993) find more than an order of magnitude larger,
13\%, in mapping $620\,\rm deg^2$ in the CO line with the CfA
$1.2\,\rm m$ telescope.  They find that the molecular surface density
of these high latitude cirrus, projected on the Galactic plane, is
between $N_{\rm H} = 0.5$ to $1\cdot 10^{20}\,\rm cm^{-2}$, which
corresponds to 20\% to 50\% of the total molecular surface density of
the Milky Way.  However, the controversy is not so large when the
limiting surface density is considered: Heithausen et al.~(1993) find
a larger surface filling factor, but they include larger and less
opaque cloud envelopes. Magnani et al.~ (1996) in compiling all
literature data confirm their filling fraction of $5\cdot 10^{-3}$,
which corresponds to $10-20\%$ of the local Milky Way molecular
surface density. The two estimations differ by a factor 2, and if we
take the overlapping value of 20\%, we end up with an average column
density of $\langle N_{\rm H} \rangle = 0.5 \cdot 10^{20}\,\rm
cm^{-2}$.  This estimation has been made with a low conversion factor
for the HLC's: $N$(H$_2$)/$I$(CO) ratio of $0.5\cdot 10^{20}\,\rm
mol\, cm^{-2}\,K^{-1}\,km^{-1}\, s$) for Heithausen et al.~(1993), and
a mixed range of values in the compilation of Magnani et al.~(1996);
this estimation is therefore highly uncertain (by up to an order of
magnitude), and is likely to be a lower limit. The low conversion
factor adopted for HLC's makes their mass much lower than the virial
mass, and their CO abundance much larger than that of local dark
clouds (e.g.~Boden \& Heithausen 1993). The standard conversion ratio
(cf.~Strong et al.~1988) would lead to $\langle N_{\rm H}({\rm HLC})
\rangle = 2.5 \cdot 10^{20}\,\rm cm^{-2}$. The value derived by Reach
et al.~(1995) for the cold component at high latitude ($|b|>
30^\circ$) corresponds to an average column density of $\langle N_{\rm
H} \rangle = 5 \cdot 10^{20}\,\rm cm^{-2}$, and is therefore
compatible with the HLC value, within the uncertainties. Moreover,
since Reach et al.~have ignored any CIBR background, their average
column density of the cold molecular material at high latitude could
be somewhat overestimated (see Fig.~4).  This overestimation of dust
optical depth at high latitudes is acknowledged by Reach et
al.~(1995): they subtracted from the FIRAS spectra the brightest CIBR
model from Franceschini et al.~(1994), before repeating their spectral
fit; the optical depths decreased, but the notable point is that the
temperatures of the warm and cold components are barely affected.

Note that the remarkable correlation between optical depths of the
cold and warm COBE components, even at high latitude, supports the
interpretation of the cold component as coming from the cold dust
shielded by the warm envelopes of the same clouds. The mean ratio 7 of
optical depths found by Reach et al.~(1995) reveals that the molecular
mass of the Galaxy could be 7 times larger than previously assumed.

Whatever the origin of the cold component is, it might be a serious
problem for measurements of cosmic microwave background fluctuations;
it is not perfectly correlated spatially with the warm dust, and has
the temperature and spectrum of the cosmological signal. Moreover,
since it is likely that this component exists also in every spiral
galaxy similar to the Milky Way, at a temperature $T_{\rm bg0} (1+z)$,
the integrated redshifted signals over all galaxies contribute also to
the microwave fluctuations.

\section{Conclusions}
It is well established from the big-bang nucleosynthesis and observed
abundances of primordial elements that the visible baryons today
represent only a small fraction of all baryons that must exist in the
Universe (cf.~recent measurements by Tytler et al.~1996).  The recent
micro-lensing experiments conducted towards the Magellanic Clouds have
revealed that Machos can account for 20\% of this dark baryons
(Aubourg et al.~1993; Alcock et al.~1996); the constraints are however
weak and inconclusive because 0\% as well as 100\% are still values
within the range allowed by the limited statistics of events and a
robust estimate of all the errors involved.

Cold molecular hydrogen is one of the most appealing candidate
(Pfenniger et al.~1994). Present observational constraints, when
properly assessed, do not rule out this hypothesis.  So we must search
for observational tests to constrain or infirm the proposition.

The main difficulty to detect the cold gas in emission is its low
temperature, close to the cosmic background temperature.  The
detection of the ``ultrafine" structure of the ortho-H$_2$ molecules
at km wavelengths raises considerable difficulties in the near future,
but could be eventually a good means to fix the $N$(H$_2$)/$I$(CO)
conversion ratio in our Galaxy.  The HD and LiH rotational lines are
or will be detectable by far-infrared satellites (ISO, FIRST,\dots),
but their excitation energy is too high to trace the bulk of the cold
gas. They could at best trace a perturbed fraction of it. In the same
vein, the hyperfine structure line of H$_2^+$ could be observed easily
from the ground.

Absorption lines detection might be the best way if the gas is indeed
very cold. Since the surface filling factor of the molecular clumps is
low ($f<1\%$), large statistics are required, but the perspectives are
far from hopeless. H$_2$ absorption in the Lyman and Werner bands has
already been identified in two damped Ly$\alpha$ systems. For a
clumpuscule falling just on the line of sight of a quasar, we expect a
damped absorption. It is then impossible to detect the quasar in the
optical or UV, but in the near-infrared.  Tracer molecules like CO in
absorption are much less promising due to the low metallicity in
addition to the low filling factor.

Finally, it is not excluded that the cold dust component detected by
COBE/FIRAS (Reach et al.~1995) is tracing the cold H$_2$ component,
limited to Galactic radii where the cold gas is still mixed with some
dust.
This cold component would be already 7 times more massive than the
usual warm dusty ISM. As predicted by Mathis et al.~(1983), the dense
H$_2$ cores of Giant Molecular Clouds, shielded from the ISRF by large
opacities, should keep an equilibrium temperature between 5 and
$7\,\rm K$. The fact that this cold medium is still observed at high
latitude is not in contradiction with observations of high-latitude
molecular clouds, given the large uncertainties in their true H$_2$
column densities. At high latitude, there could be also a contribution
of the cosmic infrared background radiation; the latter is however
negligible in front of the cold component at low latitude.

\bigskip
\begin{acknowledgements}
We are grateful to Claude Cohen-Tannoudji for enlightening discussions
about the quantum mechanical treatment of photon-atom interactions, to
Joachim Schaefer for interesting discussions about the H$_2$ molecule
and its dimers, and to Dick Tipping for advice about far-wing profiles
of molecular lines.  This work was supported by the Swiss ``Fond
National de la Recherche Scientifique''.
\end{acknowledgements}

\vskip2truecm\null

\appendix
\section {Appendix}
In this appendix, we compute the total cross section of the
interaction between photons and cold molecular hydrogen, including
absorption and scattering.

When a discrete excited level $j$ is coupled to a continuum of states,
the evolution of the system becomes irreversible. This is the case for
the spontaneous emission, where the probability to stay in the excited
level decreases exponentially with time, $\propto e^{-\Gamma_j t}$. 
This defines a half-life $\tau = 1/\Gamma_j$, where $\Gamma_j$ is
the transition probability per unit time. The finite life-time of the
levels implies that they acquire a finite width, of the order of
$\hbar \Gamma_j$; the energy of the level can then be represented by
$E_j+i\hbar \Gamma_j/2$.  A non perturbative quantum approach exists
to compute these photon-molecule interaction amplitudes
(e.g.~Weisskopf \& Wigner 1930).  The amplitude of absorption and
scattering has then a pole in $(\omega-\omega_{0j} - i
\Gamma_j/2)^{-1}$. This introduces an imaginary part, that corresponds
to absorption.  We consider only the molecule in its ground state,
absorbing a photon to reach the level $j$. To find the cross section,
we must first sum all amplitudes:
\begin{equation}
\Psi \propto \sum_j { {\langle j|H|0 \rangle {\rm e}^{-i\omega_{0j}\tau}} 
\over 
{\omega-\omega_{0j} + i\frac{\Gamma_j}{2}} },
\end{equation}
where $\langle j|H|0 \rangle$ is the matrix element of the
interaction, which can be considered here dipolar ($H= -{\vec d}.{\vec
E}$), since the photon wavelengths are assumed large with respect to
the sizes of the molecules.  The probability of transition from the
initial to final state is the product of the above amplitude by the
density of state, since a discrete level is coupled with a continuum
of states (the photon energy is continuous). The density of states is
$\rho(E) \propto E^2$, which introduces a term $\propto \omega^2$ as a
function of photon frequency $\omega$ (cf.~Cohen-Tannoudji et
al.~1988).  Since the square of the matrix element $\langle j|H|0
\rangle$ is proportional to the square of the dipole, and also to
$\omega$, the total cross-section is proportional to $\omega^3$; this
dependency is retrieved in the spontaneous emission coefficient
($A_{jk}$, and the related $\Gamma_j$); although very close to the
resonance, these coefficients are taken as constants, their $\omega$
dependency should be taken back to find the true shape of emission
lines, slightly different from the Lorentzian shape (Arnous \& Heitler
1953).  It can be shown that in the vicinity of an electronic dipolar
transition, between levels $j$ and $k$, the line profile has the
shape:
\begin{equation}
\sigma(\omega) = \frac{g_j}{g_k} { {\lambda_{jk}^2}\over{2\pi} }
 {{\left(\frac{\omega}{\omega_{jk}}\right)^3 \left(\frac{\Gamma_j}{2}\right)^2}
\over 
 {(\omega-\omega_{jk})^2 + \left(\frac{\Gamma_j}{2}\right)^2} }
\end{equation}
where the resonance occurs for $\omega=\omega_{jk}$ (with
corresponding wavelength $\lambda_{jk}$), $g_j$ and $g_k$ are the
statistical weights of the transition levels, and $\Gamma_j$ is the
natural width of the line (cf.~Power \& Zienau 1959).  The numerator
of this expression can be considered as a constant close enough to the
resonance, and the classical Lorentzian profile is retrieved.

According to the optical theorem, the imaginary part of the forward
scattering amplitude determines the total cross section for all
processes, elastic and inelastic, for a given initial state of the
photon (cf.~for instance Landau \& Lifchitz 1989). We can consider the
absorption cross-section from a stable fundamental state as equal to
the total cross section of all possible scattering processes, since in
absorbing a quantum, the molecule comes back to its fundamental state
after emission of one or several photons.

Therefore the total absorption cross-section from the fundamental
state, and in the vicinity of resonances, can be written as
(cf.~Landau \& Lifchitz 1989):
\begin{equation}
\sigma(\omega) = { {\omega}\over {\epsilon_0 c \hbar} } \sum_j {|d_{0j}|^2} 
{\left(\frac{\omega}{\omega_{0j}}\right)^3  \frac{\Gamma_j}{2} \over 
{(\omega-\omega_{0j})^2 + \left(\frac{\Gamma_j}{2}\right)^2} }  
\end{equation}
i.e.~the shape of the diffused line coincides with the natural shape
of the spontaneously emitted line.

It is convenient to introduce the dimensionless oscillator strengths
$f_{0j} = {{2 m_{\rm e}}\over {\hbar e^2} } \omega_{0j} |d_{0j}|^2$,
where $m_{\rm e}$ is the electron mass and $|d_{0j}|$ is the matrix
element of the dipole (and $g_j f_{jk} = g_k f_{kj}$).  The width of
each level is computed by summing the probabilities of spontaneous
desexcitation to every lower levels: $\Gamma_j = \sum_k A_{jk}$, and
the spontaneous rates are expressed by:
\begin{equation} 
A_{jk} = {{2r_0}\over{3c}}  f_{jk} \omega_{jk}^2,
\end{equation}
where $r_0 = e^2 / \left(4\pi \epsilon_0 m_{\rm e} c^2\right)$ is the
classical electron radius.  The total cross section can then be
expressed in terms of the Thomson section $\sigma_T={8\pi\over3} r_0^2$:
\begin{eqnarray}
\sigma(\omega) &=& {{3 \sigma_T c}\over{4 r_0}} 
\sum_j{ {\omega^4}\over{\omega_{0j}^4} } f_{0j}
{ \frac{\Gamma_j}{2} 
  \over 
  {(\omega-\omega_{0j})^2 + \left(\frac{\Gamma_j}{2}\right)^2} }  \nonumber \\
		&=& {{\sigma_T}\over{4}} 
  \sum_j{ {\omega^4}\over{\omega_{0j}^4} } f_{0j}
  { {\sum_k {f_{jk} \omega_{jk}^2}}\over{(\omega-\omega_{0j})^2 + 
    \left(\frac{\Gamma_j}{2}\right)^2}}
\end{eqnarray}

Although there does not exist a simple analytical formula valid over
the whole frequency range, the cross section is defined through
approximations in the three domains of non-resonant, resonant and free
(high frequencies).  At the low frequency limit $\omega \ll
\omega_{0j}$, the photon-molecule interaction reduces to the Rayleigh
diffusion $\propto \omega^4$:
\begin{equation}
\sigma(\omega) = \sigma_T \omega^4 \biggl[ 
\sum_j {{f_{j0}}\over{(\omega_{0j}^2 - \omega^2)}}\biggr]^2,
\end{equation}
and at high energy, the cross section reduces to that of free
electrons, $\sigma_T$, since the electron binding energy is then
negligible.  At high energy the cross section is anyway not realistic
since the formulae for the ground state ignore ionization and other
high-energy effects.

\begin{figure}
\vskip-1cm
\hskip-7.5mm
\psfig{angle=-90,figure=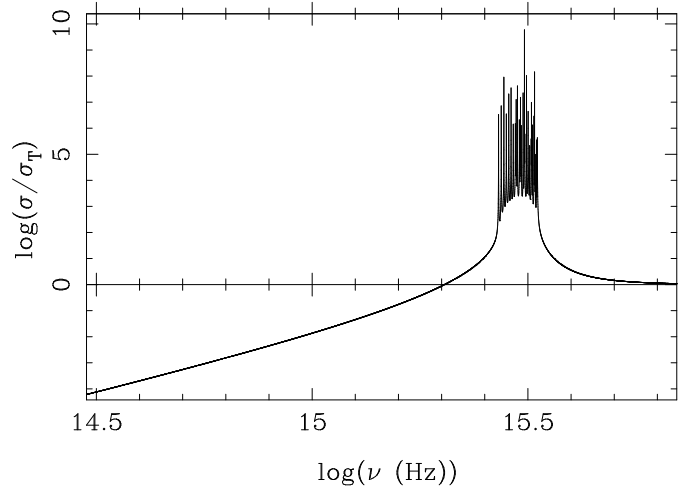,width=8cm}
\vskip-3cm
\caption[]{Computation of the cross section corresponding to the
interaction of photons with H$_2$ molecules in their ground state,
normalised to the Thomson cross section $\sigma_T = 6.652\cdot
10^{-25}\, \rm cm^2$. }
\label{scat}
\end{figure}

For the summation we consider all Lyman and Werner lines, as tabulated
in Abgrall \& Roueff (1989). Since we are concerned only with H$_2$
molecules initially in their ground state ($J=0$, $v=0$), we have
considered only $R(0)$ transitions (where $\Delta J= 1$) and tabulated
21 and 7 levels respectively in the Lyman and Werner bands.  The shape
of the total cross section is plotted as a function of frequency in
Fig.~\ref{scat}.

\vskip10truecm
\null

\end{document}